\documentclass[%
 reprint,mph,
 amsmath,amssymb,
 aps,
]{revtex4-2}
\usepackage{graphicx}
\usepackage{dcolumn}
\usepackage{bm}
\usepackage{array}
\usepackage{multirow}
\usepackage{amsmath}
\usepackage{color}
\usepackage[mathlines]{lineno}
\relax 
\begin{document}


\title{Proton Radius from  Muonic Hydrogen Spectroscopy  \\ and \\  Effect of Atomic Nucleus Motion}%

\author{Vitaly Baturin}
\altaffiliation[Corresponding author]{}
\email{baturin@jlab.org}
\address{Old Dominion University, Norfolk, VA 20052, USA}
\author{Igor~Strakovsky}
\address{The George Washington University, Washington, DC 20052, USA}



\date{\today}

\begin{abstract}
The  proton radius has been  measured in electron-proton scattering experiments and   laser based spectroscopy of muonic hydrogen. The latter method is based on the  precise  calculations for the atomic energy levels in the approximation of \textit{static nucleus}, and includes numerous corrective effects. The $4\%$ discrepancy between two 
measuring methods is known as the proton radius puzzle. We suggest that this discrepancy may be caused by an additional electromagnetic interaction with the magnetic moment generated by the \textit{nucleus motion} around the  center of mass of the muonic hydrogen. The scale of this effect is estimated based on the  known hyper fine structure of the muonic hydrogen. Our estimation show that the 
effect of the atomic nucleus motion is high enough and may help to solve the proton radius puzzle.
\end{abstract}


\maketitle


\section{Introduction}
\label{intro}

The proton charge radius  is of  fundamental importance for understanding the proton's internal structure.  From the elastic  electron-proton scattering experiments~\cite{Lehmann:1962dr,Hand:1963zz,Murphy:1974zz,Simon:1980hu,Sick:2003gm,Blunden:2005jv} and corresponding analyses~\cite{Mohr:2008fa,Pohl:2010zza}, the  average value for the proton  radius yields roughly $0.878\pm 0.011$~fm.
However, according to a more precise spectroscopy of the muonic hydrogen~\cite{Antognini:2013txn,Antognini:2013rsa} the proton radius turns out to be of $0.841\pm 0.001$~fm, 
a factor of $1.044$ lower than the value from the electron-proton scattering experiments 
This 
discrepancy is known  as the “proton radius puzzle”.

A detailed analysis of the experimental and theoretical situation around the proton radius problem, including  prospects for solutions to the  proton radius puzzle, is  given in the dedicated reviews~\cite{Antognini:2013rsa,Pohl:2013yb,Eides:2007exa}.

\section{Muonic Atom}
\label{MA:1}
According to Dirac equations  with a Coulomb  field,  the atomic shells ${2S_{}}$ and ${2P_{}}$ of the  hydrogen atom should have the same energy. The Lamb shift is the violation of this rule which is observed as electromagnetic transition between the energy levels of the aforementioned atomic shells.  

As it was shown soon after its discovery,  the Lamb shift is  caused by  the interaction of  atomic electrons  with  vacuum fluctuations such as virtual electron-positron pairs.

The modern  theory of light hydrogenic atoms based on the Dirac equation with a Coulomb~\cite{Eides:2007exa} accounts for  numerous corrections to the Lambs shift.  

A theoretical description 
of the  muonic atom  is similar to that of the ordinary hydrogen, provided the electron mass ($m_{e}$)  is replaced with  the reduced mass of muon: 
\begin{equation}
    m_{\mu}^{r} = \frac{m_{\mu}}{(1+\frac{m_{\mu}}{m_p})}=95~MeV/c^2,
    \label{eq:murma}
\end{equation}
where $m_{p}$ and $m_{\mu}$ are the masses of the atomic nucleus (proton) and orbital muon, respectively.

The energy levels of the muonic atom are determined using the precise relativistic wave functions calculated in the Coulomb field of the static nucleus. However,  there are  specific effects for the  muon hydrogen caused by that the orbital muon is $m_{\mu}/m_e = 207$ times closer to the nucleus.


In particular, due to a smaller muonic atom size, the wave-function of the orbital muon overlaps with the  atomic nucleus ${(m_{\mu}/m_e)}^3 \approx 10^{7}$ times stronger than in the regular hydrogen that makes it more sensitive to the proton size. Also one may expect a significant  sensitivity of the muonic hydrogen shells to the electromagnetic form factor of the proton, as well as to its polarization in the electric field of the muon.

As the size of virtual electron-positron pairs is compatible with the  radius of the muonic hydrogen, the last is  more sensitive to vacuum fluctuations, that makes it easier to measure the Lamb shift.  In the ordinary hydrogen the degenerated $2S_{1/2}$ and $2P_{1/2}$ 
levels split  due to the Lamb shift by $1058$~MHz ($4~\mu$eV). Since in the muonic hydrogen two  particles are very close to each other the corresponding Lamb shift is $\approx 10^5$ times  higher ($206$~meV).

 Consequently, the muonic atom shells are very sensitive to the spatial structure of the atomic nucleus and
 the Lamb shift experiment~\cite{Antognini:2013txn,Antognini:2013rsa} provides a significantly better precision for the proton radius measurement compared to the regular hydrogen spectroscopy and electron–proton scattering experiments. 

How magnetic forces affects the muonic atom levels?
The atomic levels   are classified by the orbital momentum $\vec l$, the total momentum $\vec j = \vec l + \vec s$, where $\vec s$ is the \textcolor{black}{orbital}  particle spin, and  the full moment \textcolor{black}{of the atom}  $\vec F = \vec j + \vec I$, where $\vec I$ is the spin of \textcolor{black}{the atomic} nucleus.

The \textcolor{black}{fine (spin-orbit)} interaction of magnetic moments shifts $j = 1/2$ level with respect to $j = 3/2$ state. 
Interaction of the nucleus magnetic moment with the atomic magnetic fields (hyper-fine interaction) further splits the energy levels of  $S$ and $P$ shells on two different levels depending on the orientation of the nucleus spin $\vec I$.  Hence the energy level depends on the  total angular  moment   
$\vec F$ of the atom. 
Thus, both the regular hydrogen and the muonic hydrogen atoms manifest six separated levels shown in Fig.~\ref{fig:02}. 
Since in the muonic atom particles are 207 times closer than in regular hydrogen the separation of energy levels is obviously higher for the muonic atom.

In the rigorous theory~
\cite{Adamczak:2012zz}~(p.~72, Eq.~(2))  the part of the Hamiltonian responsible for the hyper-fine interaction is proportional to the magnetic moment of the nucleus. For example, the hyper-fine splitting in the muonic hydrogen in the $S$ state, including nucleus size effects,  is proportional to the so called Fermi energy $E_{F}$: 
\begin{equation}
    E_{F} =\mu_p\frac{8\alpha^4}{3n^3} \frac{m_p^2m_{\mu}^2}{(m_p+m_{\mu})^3} \propto \mu_{p}m^2_{\mu},
    \label{eq:hf}
\end{equation}
where $m_{\mu,p}$ are the atomic particle masses defined in Eq.~(\ref{eq:murma}), $\mu_p$ is the nucleus (proton)  magnetic moment, and $n$  - principle quantum number.
With this relation we  emphasize that in the rigorous approach~\cite{Faustov:2001pn,Adamczak:2012zz} the energy of the hyper-fine interaction  is proportional to the magnetic moment of the $static$ atomic nucleus and second power of a lepton mass.

In this article, we focus on the hyper-fine interaction of the atomic muon with 
the atomic nucleus  and consider the validity of the \textit{static nucleus approximation} to the interpretation of the very precise results of the Lamb Shift Experiment. 




\section{Lamb Shift Experiment}
\label{LSE:0}

\begin{figure}[!htb]
\center{\includegraphics[width=85 mm]{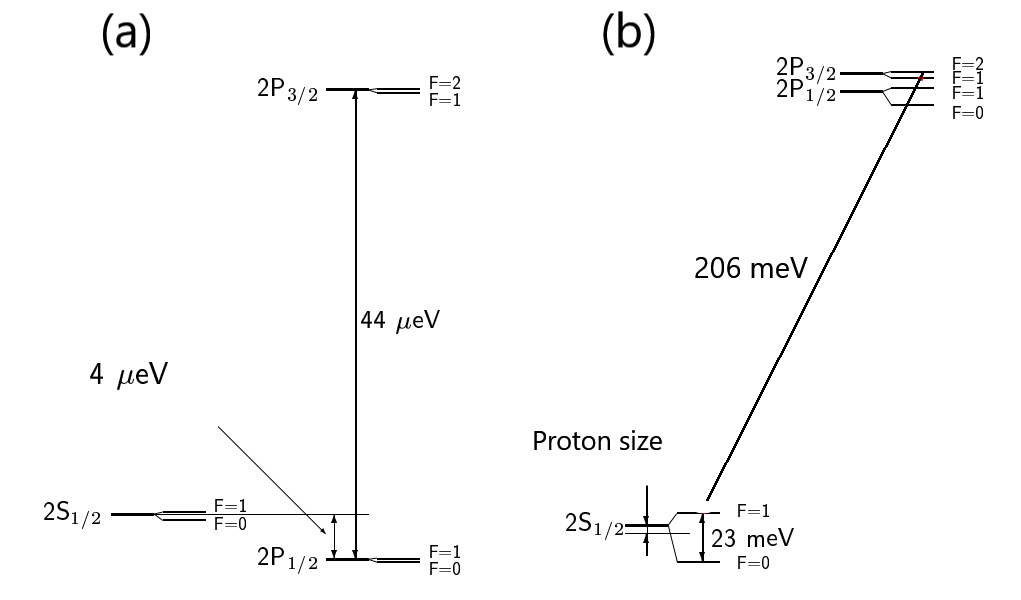}}

\caption{Hyperfine terms of hydrogen-like atoms with the principal quantum number $n = 2$. 
    (a)~Regular hydrogen. The $4~\mu$eV interval is the “classical” Lamb shift with $2S_{1/2}$ state higher than $2P_{1/2}$.
    (b)~Muonic hydrogen. The vacuum polarization pulls the $2S_{1/2}$ states below the $2P_{1/2}$ states. The $2P_{3/2}^{F=1}$-$2P_{3/2}^{F=2}$ splitting is $3.4$~meV~\cite{Pohl:2013yb}.}
\label{fig:02}
\end{figure}

The Lamb shift experiment is focused on the transition between $2S_{1/2}^{F=1}$ and $2P_{3/2}^{F=2}$ states (Fig.~\ref{fig:02}) of muonic hydrogen~\cite{Antognini:2005}, and aims to improve the precision of the proton radius measurement by a factor of $20$. The method of this experiment is as follows:

Low-energy negative muons enters the pure hydrogen-filled vessel, where they come to a rest. The target vessel is exposed to $5~T$ magnetic field used to focus the low-momentum beam.

When negative muons stop, the excited muonic hydrogen atoms are formed, a majority of which de-excite within $100$~ns to the ground state. However, about $1\%$ of muonic atoms populate a meta stable $2S_{1/2}^{F=1}$ state. At the atmospheric pressure, this shell de-populates quickly due to a very frequent collisions of muonic atom with gas molecules. 

In order to reduce the de-population rate of the meta stable state  the pressure  in the Lamb shift experiment is as low as $10^{-3}$ bar and, correspondingly, the $2S_{1/2}^{F=1}$ state lifetime is of $1.3~\mu s$. This relatively  short period is long enough to generate a trigger for the state of art laser system.  

\subsection{Laser System}
\label{LSE:2}

The function of the tunable laser system is to stimulate a  resonant transition between the aforementioned $2S_{1/2}^{F=1}$ and $2P_{3/2}^{F=2}$ muonic hydrogen levels. The proton radius is determined from the observed  resonance frequency. The laser system  is a cascade of several laser based devices with the wavelength of the final device tunable around $50$~THz. On the  muon stop  trigger,  with a rate of $10^3$~$s^{-1}$, the laser system  generates a $5$~ns light pulse with the delay of $1.5$~$\mu s$ only, that is compatible with the lifetime of the initial $2S_{1/2}^{F=1}$ state. A powerful light pulse ($0.25~mJ$) illuminates the mirrored inner  space of the  target volume that contains hydrogen with recently formed muonic atoms. If the tunable laser is on resonance with the $2S_{1/2}^{F=1}$-$2P_{3/2}^{F=2}$ transition frequency, then such transition  occurs in about $30\%$ of muonic atoms. Within a very short lifetime of $8.5$~ps, the muonic hydrogen $2P_{3/2}^{F=2}$ state de-excite to $1S_{1/2}$ ground state emitting $1.9$~keV photons. These photons  are detected by the  array of avalanche photo-diodes covering  a large area around the muon stop volume. 

\subsection{Resonance Frequency and Proton Radius}

The resonance frequency is determined by observing the yield of the $1.9$~keV photons in a function of the varying wavelength. Such dependency manifests a peak at $f_{R} = 49.885$~THz corresponding to the $2S_{1/2}^{F=1}$-$2P_{3/2}^{F=2}$ transition. 

The value for the proton radius yields from thus measured resonant frequency $f_{R}$ using the relation~\cite{Antognini:2013rsa}~ (p.~14,~Eq.~(2.18)) obtained for the Lamb shift with the precise relativistic wave functions:
\begin{equation}
    f_{R}/THz = 50.7700 - 1.2634~r_p^2 + 0.008394~r_p^3,
    \label{f=2s}
\end{equation}
where the proton radius 
is given in femtometers $(fm)$ and is defined as the
RMS radius of the spherically symmetric distribution of the proton charge density.
In terms of energy, the Lamb shift is parameterized~\cite{Antognini:2005}~ (p.~1,~Eq.~(1.1))  as:
\begin{equation}
    \Delta E_{R}/meV = 209.9685 - 5.2248~r_p^2 + 0.0347~r_p^3,	
    \label{E=2s}
\end{equation}
Here, the major contribution ($\approx 209$~meV) is given by the effects of vacuum polarization. This "pedestal" term 
also accounts for more than  20 corrections  that influence the energy levels of $2S_{1/2}^{F=1}$ or $2P_{3/2}^{F=2}$ state~\cite{Antognini:2005,Eides:2007exa}.

The second and third terms of Eq.~(\ref{E=2s})~ are sensitive to the proton radius. The most sensitive to the proton radius second term is given by the relation~\cite{Eides:2007exa}~(p.~110,~Eq.~(6.3)) or \cite{Antognini:2005}~(p.~136,~Eq.~(D.5)):
\begin{equation}
    \Delta E_2 = -\frac{2}{3} \alpha^4 n^{-3} m^3_{\mu} r_p^2\delta_{l0} 
    = -5.2248r_p^2 = -4.03~meV,
\label{eq:deltaE2}
\end{equation}
where $n = 2$ is the atomic principle quantum number of only $S$-state, since $P$-state energy level is insensitive to the nucleus size,
$m_{\mu}$ - the reduced mass of the muon, $\alpha$ - Sommerfeld's constant, and $r_p=0.878~fm$ is used for the numerical estimate.

\section{Advantages and Questions}
\label{AAH:0}

Although according to  Eq.~(\ref{eq:deltaE2}), the effect of proton size in the muonic $2S_{1/2}^{F=1}$-$2P_{3/2}^{F=2}$ Lamb shift is quite small, it is about hundred times higher than in the ordinary hydrogen. That’s what makes the muonic hydrogen spectroscopy a very precise and sensitive measuring instrument. 

 
 
 However, the basic relationships for the proton radius Eqs.~(\ref{f=2s})~ and ~(\ref{E=2s})~  include numerous contributions, and there is no guarantee that there are no other effects that can alter the interpretation of the measured frequencies and the corresponding transition energies.
 

\subsection{Nucleus Spin}

The proton radius yields from the Eq.~(\ref{E=2s})~ which is based on the precise relativistic wave functions calculated in a static Coulomb field. 
It is important to notice  that the Dirac equation with a static Coulomb field accounts for  the  spin of electrons and muons,  but  ignores the spin  of the atomic nucleus. Therefore, the interaction of the  atomic magnetic moment  with the  magnetic moment of proton is accounted separately~\cite{Antognini:2005}.
In particular, according to the precise calculations performed for the  $2P_{3/2}^{F=2}$  state, the interaction  with nucleus spin contributes as much as $1.3$~meV~\cite{Antognini:2005} to the $2S_{1/2}^{F=1}$-$2P_{3/2}^{F=2}$  energy interval.

\subsection{Static Nucleus Approximation}  
\label{efforb}
From the relevant reviews~\cite{Carlson:2015jba,Antognini:2015vxo}, we know that the energy levels of muonic hydrogen are calculated using the wave functions obtained in the approximation of an infinitely heavy stationary nucleus.  

However, the  recoil corrections are taken into account and summarized in Table~2.2~\cite{Antognini:2005}. From this Table the total recoil effect may be estimated as 0.0705 $meV$, or $1.7\%$ of the proton radius effect (Eq.~(\ref{eq:deltaE2})).

Although the influence of the finite mass of the nucleus is taken into account through the reduced muon mass and the nucleus recoil momentum~\cite{Antognini:2005,Karshenboim:2012wv}, the static nucleus approximation may not correspond to reality. 

The reason is that  both particles are actually "orbiting" the common center of mass. Hence the  atomic nucleus can generate an additional  magnetic moment which is referred below as the "induced magnetic moment". As a result, the Coulomb potential which has been used for precise  calculations of atomic $S$- and $P$-levels, has to be modified accordingly. 

\section{Effect of Orbiting Nucleus}
As it is known from the atomic physics the magnetic moment of  the orbital electron may be presented as 

\begin{equation}
    \mu_{e} = \frac{e\hbar}{2m_ec}\sqrt{l(l+1)}=\mu_{B} \sqrt{l(l+1)},
    \label{eq:pmag}    
\end{equation}
where $l$ is  the  orbital moment of electron, $\mu_B$ stands for the Bohr magneton.
In the muonic atom  the muon mass should be used on  place of the electron mass.

What may be the magnetic moment of an orbiting nucleus? 
Let's first estimate  it in classical approximation, where one may  consider a hydrogen-like atom as  two charged particles rotating around the common center of mass  with the same angular frequency $\omega$. The absolute values of the orbital angular moments $l_{\mu,p}$ and magnetic moments $\mu_{\mu,p}$ for each  particle  are:
\begin{align}
\l_{\mu,p} = m_{\mu,p}\omega r_{\mu,p}^2, \nonumber \\ \mu_{\mu} =  -\frac{e}{2m_{\mu}}l_{\mu} = -\frac{e}{2} \omega r_{\mu}^2, \nonumber \\
\mu_{p} =  \frac{e}{2m_{p}}l_p = \frac{e}{2} \omega r_{p}^2,
\label{eq:pmag1}     
\end{align}
where $\omega$ is the angular  frequency, $m_{\mu,p}$ are the masses of muon and proton, respectively, and $r_{\mu,p}$ are  corresponding  particle orbit radii.
Since both particles "orbit"  the common center of mass, the rotation periods  are identical and the obvious relation holds: $m_{\mu}r_{\mu}=m_{p}r_{p}$. 
For convenience, using Eq.~(\ref{eq:pmag1}),  we express  the orbital  magnetic moment of the  
proton $\mu_p$  in terms of  the magnetic moment of the orbiting partner, i.e., muon:    
\begin{equation}
    \mu_{p} =  \mu_{\mu}\bigg( \frac{m_{\mu}}{m_p} \bigg)^2.
    \label{eq:mupmumu}
\end{equation}
It is interesting to notice that although the total orbital  moment $l_p+l_{\mu}$ coincides with the orbital moment calculated for the muon reduced mass orbiting the motionless nucleus, the effective magnetic moment of such atom does not follow this rule, and  in the limit of equal masses the effective magnetic moment  zeroes.  

In accordance to Eq.~(\ref{eq:pmag}) the smallest nonzero value for $ \mu_{\mu}$ is  $\sqrt{2}\mu^{(\mu)}  $  where $\mu^{(\mu)}$ $= 4.5$ $\times$ $10^{-26}$ $JT^{-1}$  is  the nominal magnetic moment of muon.


\subsection{Resonance Shift}

Fortunately, the estimation of the induced magnetic moment  effect in the  $2S_{1/2}^{F=1}-2P_{3/2}^{F=2}$  resonant transition may be done without going into detailed theoretical calculations. With such goal in mind,  we focus on the $2P_{3/2}^{F=2}$ and $2P_{3/2}^{F=1}$ states of a muonic atom, which have  opposite  orientations of the  nucleus spin (Fig.~\ref{fig:02}).

Precision calculations for static nucleus show (\cite{Pohl:2013yb},~p.~27, Eq.~(32) and ~p.~60,~Fig.~4) 
that interaction of the nucleus magnetic moment  with the atomic magnetic field results in the following  hyper fine 
splitting of $2P_{3/2}^{F=2}$ and $2P_{3/2}^{F=1}$ states: 
\begin{equation}
    {\Delta E}= 
    (3.3926- 0.1446)~meV\approx 3.25~meV, 
    \label{eq:epsil}
\end{equation}
which is obviously proportional to the doubled value of the nominal magnetic moment $\mu_{N}$ of the nucleus (proton) via Fermi energy, as prescribed by  Eq.~(\ref{eq:hf}) for all hyper fine effects.  

Since the "radius" of the proton orbit is very small compared to the atomic radius, we reasonably assume that the orbital proton interacts with the same magnetic field as  in the center of the atom.  Hence, the effective  magnetic moment may be considered as a vector sum of the nucleus nominal magnetic moment and its orbital magnetic moment.

Correspondingly the effect of the induced magnetic moment may be considered as a small linear perturbation of the value ${\Delta E}$ from Eq.~(\ref{eq:epsil}). 
In order to estimate the  effect of the nucleus  orbital motion we  scale a halve of  this  precise value   by the ratio of the induced magnetic moment of the orbital proton (Eq.~(\ref{eq:mupmumu})) to its nominal magnetic moment $\mu^{(p)} = 1.41\times10^{-26}JT^{-1}$ used in Eq.~(\ref{eq:epsil})~:
\begin{equation}
    \Delta W \approx\frac{\mu_p}{2\mu^{(p)}}{\Delta E} 
    = \sqrt{2}\frac{\mu^{(\mu)}}{2\mu^{(p)}}\bigg(\frac{m^r_{\mu}}{m_p}\bigg)^2{\Delta E} 
    \approx 0.1~meV.
    \label{eq:wfracmu}
\end{equation}
Note that the magnetic moment of the orbital  proton is opposite to that of negative muon, and similar to the static nucleus,  the induced magnetic moment shifts the $2P_{3/2}$  state to higher energies, as well as the transition energy, leading to a higher proton radius, provided the measured resonant transition energy $\Delta E_R$  in  Eq.~(\ref{E=2s})~ remains unchanged. 
The estimate Eq.~(\ref{eq:wfracmu})~  constitutes  $2.5\%$  of the  energy specified in Eq.~(\ref{eq:deltaE2}), that correspondingly  translates  to the $\approx1\%$ effect in terms of proton  radius.

\subsection{Orbital Proton and Bohr Magneton}

The estimate Eq.~(\ref{eq:wfracmu})~ is based on Eq.~(\ref{eq:mupmumu}), which suggests  a direct proportionality of the induced magnetic moment to the second power of the proton orbit radius. This relation is based on the assumption of classical behaviour of atomic components, i.e., that there is a strict correlation between "coordinates" of two atomic particles.

In quantum case one may attribute the "orbital" nucleus with the magnetic moment defined by Eq.~(\ref{eq:pmag})~ where the mass of electron is replaced with the proton mass and  magnetic moment of the orbital particle is inversely proportional to the first power of its mass. 

Hence, following the mass dependence of the orbital magnetic moment in quantum case,  one may   estimate the orbital magnetic moment  of proton in the muonic hydrogen as: 
\begin{equation}
    \mu_p = \mu^{(\mu)}\bigg(\frac{m^r_{\mu}}{m_p}\bigg).
    \label{eq:mupbohr}
\end{equation}

As we have an additional magnetic moment,  one may expect  that atomic levels with given $F$ are split in accordance to the mutual orientation of the nucleus spin and its orbital moment. For the  projection of the resulting magnetic moment $\vec {\mu_{t}}$ to the direction of the magnetic field in the center of the atom one may write
\begin{equation}
\mu_{t}=\mu^{(p)} \pm  \mu_p=  \mu^{(p)}\bigg(1 \pm \frac{\mu^{(\mu)}}{\mu^{(p)}}\bigg(\frac{m^r_{\mu}}{m_p}\bigg)\bigg),~~or~~~\mu^{(p)}.
     \label{eq:mutot}
\end{equation}
Following the  arguments from the classical consideration above one may expect that the resulting projection $\mu_{t}$ will be higher 
than the nominal magnetic moment $\mu^{(p)}$  and, correspondingly, the $2P^{F=2}_{3/2}$ energy level  may be  higher by the value
\begin{equation}
    \Delta W \approx \frac{\mu_p}{2\mu^{(p)}}{\Delta E} 
    = \frac{\mu^{(\mu)}}{2\mu^{(p)}}\bigg(\frac{m^r_{\mu}}{m_p}\bigg){\Delta E} \approx 0.5~meV.
    \label{eq:wfr23}
\end{equation}
Note that this value has to be added to the right side of Eq.~(\ref{E=2s})~ and compared  to $4.03$~meV from the relation Eq.~(\ref{eq:deltaE2}). 
Comparing  the  estimation Eq.~(\ref{eq:wfr23})~ with the value  from   Eq.~(\ref{eq:deltaE2})~  one may conclude that the additional shift $0.5$~meV translates to $6\%$ higher proton radius ($r_p=0.89$~fm), provided that experimental value  $f_R$ in the left side of Eq.~(\ref{E=2s})~ remains as is. The last value for the proton radius almost fits the experimental value interval  $(0.867, 0.889$~fm), that was estimated from  various scattering experiments.\\

What wold be the effect of the nucleus motion in $P$-state of the ordinary hydrogen? In this relation it is interesting to notice that  from the  measurement of the $1S-3S$ transition  in regular hydrogen~\cite{Fleurbaey:2018fih} the proton radius yields as high as $0.877$~fm, while from the $2S-4P$ transition frequency, measured in Garching~\cite{Beyer:2017gug} and $2S-2P$~\cite{Bezginov:2019mdi}, it yields a lower values - $0.835$ and $0.833$~fm, respectively.

\section{Conclusion}

We have considered  the effect of the nucleus orbital motion around the  atom's center of mass as a perturbation to the accurately calculated  data on the muonic hydrogen hyper-fine structure with static nucleus. We have shown  that nucleus motion is a possible source for the atomic levels distortion, leading to  the underestimation of the proton radius. Hence, the  interaction of the orbital  atomic nucleus  with the atomic magnetic field needs to  be strictly accounted when calculating the transition energies. 


The effect of nucleus motion may  significantly change the estimate of  proton radius from the Lambs shift experiment to higher values, thereby reducing, or even eliminating, the discrepancy that makes up the proton radius puzzle.

\section*{Acknowledgements}
The work of I.S. was supported in part by the U.S. Department of Energy, Office of Science, Office of Nuclear Physics, under Award No.~DE–-SC0016583.





\end{document}